\documentclass[preprint,aip]{revtex4-2}

\usepackage[utf8]{inputenc}
\usepackage{xprintlen}
\usepackage{amsmath}
\usepackage{physics}
\usepackage{mhchem}
\usepackage{url}
\usepackage{amsfonts}
\usepackage{amssymb}
\usepackage{tabularx}
\usepackage{mathtools}
\usepackage{subfigure}
\usepackage{booktabs}
\usepackage{makeidx}
\usepackage{graphicx}
\usepackage{lmodern}
\usepackage{siunitx}
\usepackage{upgreek}
\usepackage{adjustbox}
\usepackage{blindtext}
\usepackage{siunitx}
\usepackage[section]{placeins}
\usepackage{float}
\usepackage{braket}
\usepackage{appendix}
\usepackage{xr}
\begin{document}

\title{Laser stimulation of muscle activity with simultaneous detection using a diamond colour centre biosensor}
\author{Luca Troise}
\affiliation{Center for Macroscopic Quantum States (bigQ), Department of Physics, Technical University of Denmark, Kgs. Lyngby, Denmark}%
\author{Nikolaj Winther Hansen}%
\affiliation{Department of Neuroscience, University of Copenhagen, Copenhagen, Denmark}%
\author{Christoffer Olsson}
\affiliation{Department of Health Technology, Technical University of Denmark, 2800 Kgs. Lyngby, Denmark}%
\author{James Luke Webb}
\affiliation{Center for Macroscopic Quantum States (bigQ), Department of Physics, Technical University of Denmark, Kgs. Lyngby, Denmark}%
\thanks{Main Corresponding Author}
\email{jaluwe@fysik.dtu.dk}
\author{Leo Tomasevic}
\affiliation{Danish Research Centre for Magnetic Resonance, Centre for Functional and Diagnostic Imaging and Research, Copenhagen University Hospital - Amager and Hvidovre, Kettegaard Alle 30, 2650 Hvidovre, Denmark}%
\author{Jocelyn Achard}
\affiliation{Laboratoire des Sciences des Proc\'ed\'es et des Mat\'eriaux, Universit\'e Sorbonne Paris Nord,  93430 Villetaneuse, France}%
\author{Ovidiu Brinza}
\affiliation{Laboratoire des Sciences des Proc\'ed\'es et des Mat\'eriaux, Universit\'e Sorbonne Paris Nord,  93430 Villetaneuse, France}%
\author{Robert Staacke}
\affiliation{Division Applied Quantum System, Felix Bloch Institute for Solid State Physics, Leipzig University, 04103, Leipzig, Germany}%
\author{Michael Kieschnick}
\affiliation{Division Applied Quantum System, Felix Bloch Institute for Solid State Physics, Leipzig University, 04103, Leipzig, Germany}%
\author{Jan Meijer}
\affiliation{Division Applied Quantum System, Felix Bloch Institute for Solid State Physics, Leipzig University, 04103, Leipzig, Germany}%
\author{Axel Thielscher}
\affiliation{Department of Neurology, Copenhagen University Hospital Bispebjerg and Frederiksberg, Bispebjerg Bakke 23, 2400 Copenhagen, Denmark}%
\affiliation{Department of Health Technology, Technical University of Denmark, 2800 Kgs. Lyngby, Denmark}
\author{Hartwig Roman Siebner}
\affiliation{Department of Clinical Medicine, Faculty of Health and Medical Sciences, University of Copenhagen, Blegdamsvej 3B, 2200 Copenhagen N, Denmark}
\affiliation{Danish Research Centre for Magnetic Resonance, Centre for Functional and Diagnostic Imaging and Research, Copenhagen University Hospital - Amager and Hvidovre, Kettegaard Alle 30, 2650 Hvidovre, Denmark}%
\author{Kirstine Berg-S{\o}rensen}
\affiliation{Department of Health Technology, Technical University of Denmark, 2800 Kgs. Lyngby, Denmark}
\author{Jean-Fran\c{c}ois Perrier}
\affiliation{Department of Neuroscience, University of Copenhagen, Copenhagen, Denmark}
\author{Alexander Huck}
\affiliation{Center for Macroscopic Quantum States (bigQ), Department of Physics, Technical University of Denmark, Kgs. Lyngby, Denmark}
\email{alexander.huck@fysik.dtu.dk}
\author{Ulrik Lund Andersen}
\affiliation{Center for Macroscopic Quantum States (bigQ), Department of Physics, Technical University of Denmark, Kgs. Lyngby, Denmark}
\email{ulrik.andersen@fysik.dtu.dk}

\begin{abstract}
\textbf{Abstract}: The detection of physiological activity at the microscopic level is key for understanding the function of biosystems and relating this to physical structure. Current sensing methods often rely on invasive probes to stimulate and detect activity, bearing the risk of inducing damage in the target system. In recent years, a new type of biosensor based on color centers in diamond offers the possibility to passively, noninvasively sense and image living biological systems. Here, we use such a sensor for the \textit{in-vitro} recording of the local magnetic field generated by tightly focused, high intensity pulsed laser optogenetic neuromuscular stimulation of the extensor digitorum longus muscles. Recordings captured a compound action potential response and a slow signal component which we seek to explain using a detailed model of the biological system. We show that our sensor is capable of recording localized neuromuscular activity from the laser stimulation site without photovoltaic or fluorescence artifacts associated with alternative techniques. Our work represents an important step towards selective induction of localized neurobiological activity while performing passive sensing and imaging with diamond sensors, motivating further research into mapping of neural activity and intra-cellular processes.
\end{abstract}

\maketitle

\section{Introduction}

Sensing biological activity at the microscopic level is of fundamental importance for understanding the basic processes in living organisms \citep{MCDONALD2012495}. Relating biological activity to cell structure and to the resulting behavior are key steps to decipher biological functions in health and their alterations in disease. This can inform new diagnosis, for example of early stages of neurodegenerative disorders, including peripheral nerve damage and muscle atrophy. Of particular interest is the biological response to a stimulus, such as a pulse of electrical current or light, and being able to detect this response with high spatial and temporal resolution \citep{Scanzani2009}.  Many different biosensor approaches have been developed for microscopic direct detection of activity, in particular recording of electrical activity using probe electrodes (electrophysiology) and by sensing using fluorescent biomarkers \citep{GRIENBERGER2012862,Lin2016-us}. However, such methods are highly invasive, particularly when recording deep within tissue. This entails the risk of tissue damage, through mechanical action, toxicity or infection, which could be detrimental for living organisms and might adversely affect the sought after activity \citep{GECHEV201662,10.3389/fbioe.2020.00416}.

An alternative method is to indirectly detect activity, passively using a sensor outside of the sample or subject. For electrical activity in the brain or nervous system, this can be achieved by monitoring the magnetic field induced by the signal, which can freely penetrate biological tissue. Field sensing has been demonstrated for larger scale tissue sensing in established techniques such as magnetoencephalography (MEG) and magnetocardiography (MCG), using superconducting quantum interference device (SQUID) sensors \citep{Clarke2018, Hmlinen1993}. However, these sensors have significant disadvantages, requiring special shielding from background magnetic field (e.g. geomagnetic, mains electricity) and cryogenic cooling, limiting these valuable techniques to very few facilities and applications. 

In recent years, new biosensor approaches without these drawbacks have been actively pursued \citep{arlett2011,baselt1998,Boto2018}. In this work, we focus on an approach utilising colour centres, defect sites in a solid state material with optical properties highly sensitive to their physical parameters and local environment. In particular, interest has focused on the negatively charged nitrogen vacancy (NV) centres in diamond, consisting of a nitrogen subsitutional dopant paired with a lattice vacancy. NV centres have been used for sensing of magnetic field \citep{Taylor2008}, electric field \citep{Dolde2011}, temperature \citep{Neumann2013} and strain (pressure/force) \citep{Kehayias2019} via optically detected magnetic resonance (ODMR) spectroscopy \citep{odmr2009,doi:10.1063/1.3385689}. Diamond NV sensors offer advantages including high sensitivity \citep{Fescenko2020,Wolf2015}, room temperature operation and high spatial \citep{Mizuno2020, LeSage2013} and temporal \citep{Barry2020} resolution. Diamond is highly biocompatible, enabling biosensors based on NV centres to work in solution across a wide range of temperatures and pH values \citep{toyli2012}. Such sensing can be performed remotely at a distance or in close contact (and even within) a biological specimen \citep{Barry2016, Wu2021-ee}.

Sensors based on NV centres in diamond are particularly suited to microscopy applications in either widefield or confocal configurations \citep{LeSage2013,PhysRevApplied.10.044039,doi:10.1126/science.276.5321.2012}, in a manner SQUIDS or other alternative sensors cannot easily achieve. NV sensors can sense a small volume region, containing single (neuron) cells and down to the few or single molecule level. A desirable goal is to selectively and precisely optically stimulate activity while recording activity close to or at the site of stimulation, without suffering from the measurement artifacts that pose challenges for existing sensors \citep{Kozai2015-om, Cardin2010-tr,Packer2013-vd}. Achieving this goal is key to realising applications including nanoscale nuclear magnetic resonance \citep{Staudacher2013-ye}, radical pair sensing for avian magnetoreception \citep{Xu2021-ab}, intracellular nanodiamond studies \citep{Kucsko2013, Fujiwara2020,McGuinness2011} and stimulation of neurons in brain tissue \citep{Hall2012, Karadas2018,Price2020}. Due to the confinement required, focused laser stimulation is preferable for these applications, with the higher laser optical intensity ideal for for stimulation well within tissue. 

In this work, we seek to demonstrate the viability of such focused laser stimulation combined with simultaneous microscopic sensing from biological tissue using a diamond NV sensor, located in proximity to the stimulation site. We use living \textit{extensor digitorum longus} (EDL) muscles from mice \textit{in vitro}, genetically modified to contain the light gated cation channel channelrhodopsin (ChR2) \citep{Webb2021}. Distinct from our previous work, we stimulate the muscle tissue using focused pulsed laser light, of a specific wavelength and intensity, generating highly localised ($\approx$150$\mu$m) stimulation and activation of the muscle response. We then record the response of the muscle via the magnetic field induced by ionic current associated with compound action potentials in the muscle.  We seek to 1) demonstrate highly focused stimulation with artifact-free recording of activity using our diamond NV sensor with high temporal resolution, 2) explore the biological response recorded under laser stimulation, 3) derive new insight into the behavior of the biological system and response via the use of computer modelling and 4) employ new methodology including spectral whitening filtering of background noise to enhance biosensor signal recovery in an ordinary lab environment without magnetic shielding. 
 
\section{Material and Methods}

\begin{figure}[!h]
\begin{centering}
\includegraphics[width=164mm]{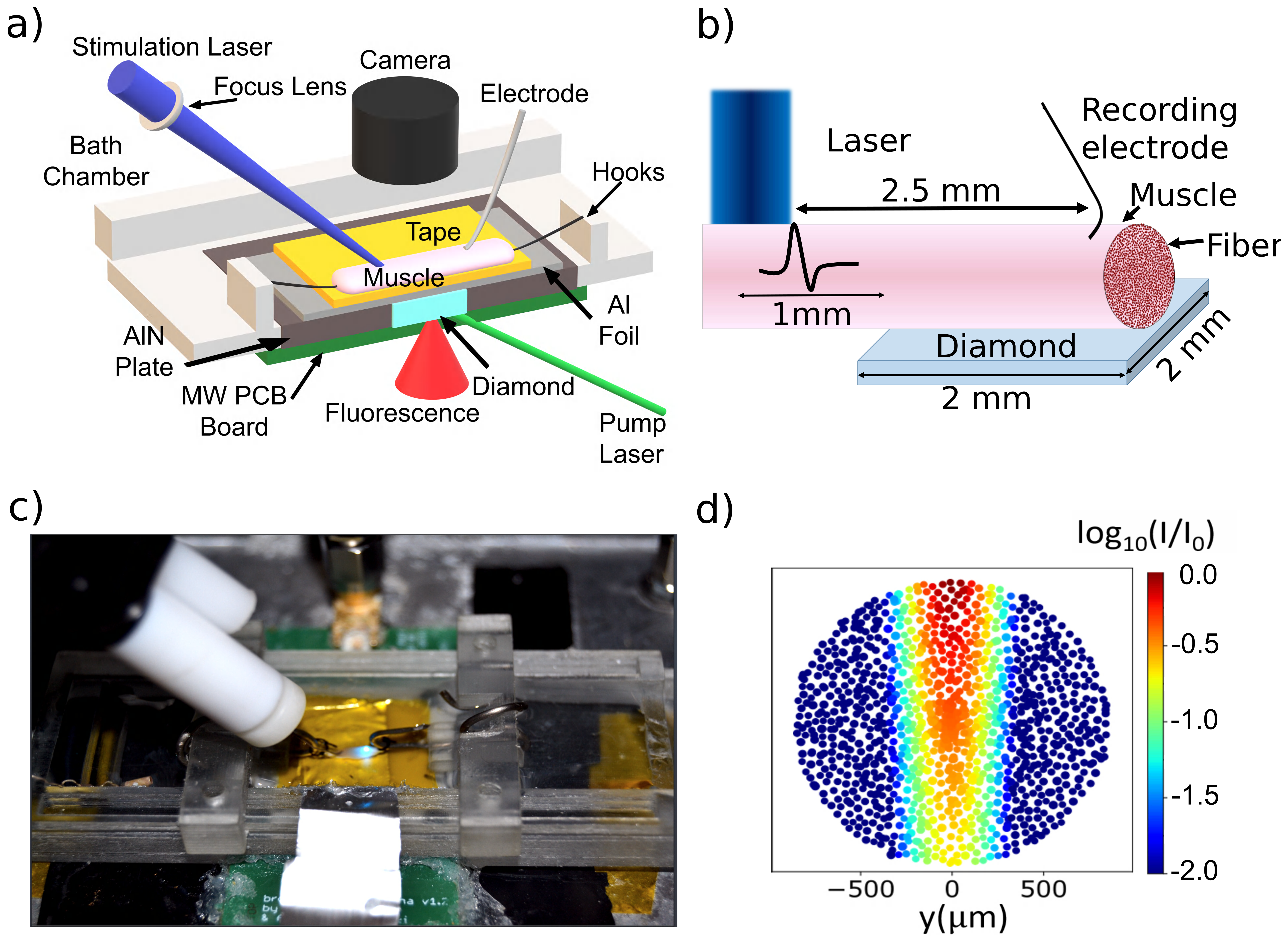}
\caption{\textbf{Experimental setup and muscle cross section used for modelling} a) Schematic of the experimental setup and b) picture of the muscle in the solution chamber with a 488nm blue stimulation laser spot incident on the muscle surface. The laser spot could be placed with high precision anywhere within the sample chamber in a 2x2cm area using mechanical mirror mounts. c) Schematic illustration of experimental setup, as used in the modelling. d)  Simulated cross-section of muscle showing the light intensity ($I$/$I_0$) distribution on the muscle fibers with laser stimulation. }
\label{Fig1}
\end{centering}
\end{figure} 

We performed \textit{in vitro} studies of EDL muscles dissected from mice, expressing the light sensitive cation channel Channelrhodopsin2 (ChR2) in muscle cells to achieve optogenetic triggering of electrical activity in the muscle using blue light. Muscles were placed in a solution chamber directly above our diamond NV sensor (Figure \ref{Fig1}) and stimulated by a 488nm laser diode in free space, focused using a long working distance lens into the sample chamber. The electrical response (\textit{electric data} in this work) was measured using an AgCl electrode in light surface contact with the muscle (Figure \ref{Fig1}) and the induced magnetic field recorded simultaneously by the diamond NV sensor (\textit{magnetic data} below). 

\subsection{Diamond NV Sensor}

We used a [100] oriented electronic-grade diamond (Element Six), of dimensions 2 $\times$ 2 $\times$ 0.5 mm$^3$, overgrown via chemical vapour deposition (CVD) with a 20$\mu$m thick diamond layer with a doping concentration of $\approx$5 ppm of $^{14}$N. NV centres were generated by 2.8MeV proton irradiation and subsequent annealing at 800\textdegree C. We measured an ODMR resonance linewidth of 1 MHz and a resonance contrast of 1.5\%. The diamond was mounted into an aluminum nitride plate heatsink, with the top surface of the diamond covered by a 16$\mu$m thick layer of aluminium foil, reflecting pump laser back into the diamond and blocking the stimulation laser light from the diamond (Figure \ref{Fig1}). The aluminium foil was covered by Kapton tape (50$\mu$m) to electrically insulate the diamond from the biological sample. The AlN plate holding the diamond was mounted on a printed board microwave antenna. On top of the diamond, we mounted a 3D printed plastic chamber for holding a solution bath, sealed using aquarium silicone.

\subsection{Sensor Operation}

We optically pumped our diamond with 1.4W of 532nm green laser light (Coherent Verdi G2), linearly polarised and coupled to the diamond from beneath at Brewster's angle (67\textdegree). Red fluorescence from the NV centres was collected using a 12mm diameter condenser lens (Thorlabs ACL 1210) via an optical filter (FEL0600). Magnetic field at the diamond was recorded as a modulation of the intensity of the red fluorescence emission recorded using an auto-balanced optical receiver (Nirvana 2007, New Focus Inc.). We measured only the magnetic field response in the direction perpendicular to the electrical current propagation direction inside the muscle, with this single axis response maximised by applying a static bias field of 1.5 mT parallel to the diamond [110] crystallographic direction. A continuous wave scheme was implemented with three-frequency microwave driving scheme \citep{ElElla2017} using two microwave generators (Stanford SG394) to drive the triplet ground state transition (2.7-3GHz) and the $^{14}$N hyperfine transitions (2.16 MHz). The two RF signals were delivered to the diamond via a printed circuit board nearfield antenna. The microwaves were modulated at 23.3 kHz for lock-in detection (Stanford SR850), using a time constant of 30$\mu$s giving a sensor measurement bandwidth of 4.8kHz. The output of the lock-in amplifier was digitised at 80 kSa/s using an analog to digital converter (NI PCI-6221) and recorded using custom-written software. 

\subsection{Muscle Preparation}

We used mice expressing the light sensitive cation channel Channelrhodopsin2 (ChR2) in muscle cells expressing Parvalbumin (PV) by crossing Gt(ROSA)26Sor\textsuperscript{tm32(CAG-COP4*H134R/EYFP)Hze}  (stock no.: 024109; Jackson Laboratory) mice with Pvalb\textsuperscript{tm1(cre)Arbr} mice (stock no.: 017320; Jackson Laboratory) \citep{Webb2021}. Adult PV-Cre::ChR2 mice were dissected as described previously (Webb et al., 2021). Quickly following euthanasia by cervical dislocation, EDL muscles were dissected in carbogen (95\% O2/5\% CO2) saturated ice-cold artificial cerebrospinal fluid (ACSF) containing (in mM): NaCl (111), NaHCO3 (25), glucose (11), KCl (3), CaCl2 (2.5), MgCl2 (1.3) and KH2PO4 (1.1). Small suture loops were tied on proximal and distal tendons. The dissected mouse muscle was kept in a solution bath of artificial cerebrospinal fluid (ACSF), chilled by passing a recycled feed of ACSF into the solution chamber through tubing submerged in ice to maintain a bath temperature between 18\textdegree C and 20\textdegree C. The ACSF was bubbled with carbogen gas to oxygenate the muscle and allow survival for up to 20 hours of measurement.     

\subsection{Laser Stimulation and Recording}

The muscle was positioned on the diamond NV sensor by two adjustable hooks and a probe electrode (AgCl coated silver wire) mounted on a micromanipulator placed in contact with the muscle top surface. The electric data from this probe electrode was recorded via a differential amplifier (Axon Cyberamp 320) relative to the grounded solution bath. Muscles were stimulated to produce compound action potentials (AP) using a TTL triggered 488nm blue laser (OdicForce Ltd.). Laser light was directed through free space to the muscle and focused onto the muscle using an f=400 mm plano-convex lens, giving a 1/e$^{2}$ laser spot size of $\approx$150$\mu$m, measured using a microscope camera mounted above the sample chamber. This beam waist could be further reduced by additional lenses in the beam path as required. Using optomechanical mirror mounts, the laser spot could be moved at sub-millimeter precision to any position on the muscle, with a travel of 2 $\times$ 2 cm$^2$ across the chamber. This represents an improvement over a confocal/inverted widefield microscopy configuration where stimulation is limited only to the centre of the objective field of view \citep{Mrzek2015,Bernardi2020}. We positioned the stimulation laser spot approximately 0.5 mm (Figure \ref{Fig1},b) along the muscle away from the diamond, to ensure we recorded signal propagation in only one direction (towards one end of the muscle). Control measurements were taken of the photovoltaic effect from laser stimulation incident to the probe electrode (see Supplementary Information, SI), and for laser stimulation directly on the diamond NV sensor. Here, no detectable artifact was observed in the magnetic data, measuring using 5ms pulses at 50mW directly onto the Kapton layer above the diamond for up to 10 hours. 

To ensure the stray field from the high current at laser turn-on was not recorded, the blue laser head was placed 3.5m away from the diamond NV sensor and the TTL trigger optically decoupled.  Optimal blue laser power was found by initially stimulating the muscle with low input laser power (5 mW), and then slowly increasing the power until a maximal response of the muscle was observed in the electric data. Optical stimulation was performed at a frequency of 0.5 Hz and with a pulse length of 5 ms. For this length, stimulation power was found to be in the range of 20-50 mW, varying between different muscles. Signals were recorded using 60s continuous time traces, giving the high spectral resolution necessary to filter and remove background magnetic noise. Many hours of recording could be acquired, limited by muscle lifespan in the solution bath (up to 20h). 

\subsection{Filtering by Spectral Whitening}

Without magnetic shielding, it was necessary to remove a substantial amount of magnetic noise from our sensor readout, including stationary mains noise at harmonics of 50Hz and non-stationary noise from building pumps and compressors. In our previous work, we demonstrated filtering using windowed notch filtering \citep{Webb2020}. However this noise removal was suboptimal, due to the use of serially generated fixed frequency windows and a single global threshold value to identify noise peaks, constraints imposed by computational time. Here, we instead implemented a Linear Time-Invariant (LTI) spectral whitening filter. This method has been used for applications including radar and detection of gravitational waves \citep{Gabbard2018,Chatziioannou2019, Roman2000}. To implement the filter, the power spectral density (PSD) of each 60s time trace was used to estimate the noise spectrum and to derive the transfer function for whitening the signals. The double-sided PSD was calculated with Welch's periodogram method with a segment length of 3.5s and 50\% overlap. The input signals were whitened in the range from 20Hz-40kHz; the lower frequency bound was chosen in order to avoid high-pass filter artifacts, the high frequency bound was set by the digitization Nyquist frequency. Following whitening, an additional 3rd order Butterworth 650Hz lowpass filter was applied. Baseline wander was removed from the trials using robust detrending \citep{de2018robust}. Further details of the filter method and optimisation are supplied in the SI. 

\subsection{Simulation and Modelling Method}

In order to verify and analyze the detected signal we performed a numerical simulation of the biosensor and muscle biosystem. We use the NEURON package \citep{Hines2001} to simulate the signal propagation in a muscle. The ion channel mechanisms were modified in accordance with a previous muscle model on EDL muscles in mice \citep{Cannon1993} (full details in SI). The optogenetic stimulation current was modelled with the use of a biophysical model developed by Foutz el al. \citep{Foutz2012} and based on the four-state model described by Nikolic et al. \citep{Nikolic2009}.

We modelled the individual muscle fibers in a cylindrical geometry with a cylinder radius of 845 $\mu$m and an individual fiber radius given by a uniform distribution of 22.5 $\pm$ 3.5 $\mu$m \citep{Augusto2017}. The fibers started at the back end of the laser stimulation spot and extended straight across the laser beam, and over across the NV diamond (Figure \ref{Fig1},b and c). The light distribution inside the muscle was calculated according to the Kubelka-Munk model. The diamond was divided into 20x20 pixels with side-lengths of 100$\mu$m, with magnetic field calculated at each pixel, and then averaged across the whole diamond. The local field potential (LFP) was calculated as a monopolar readout approximately at the site of the electrical recording tip (Figure \ref{Fig1},c). The magnetic and electrical field calculation were based on calculations performed in \citep{Karadas2018}, using the LFPy software package \citep{Linden2014}.

\section{Results and Discussion}

\subsection{Movement Inhibition}

\begin{figure}[!h]
\centering
\includegraphics[width=164mm]{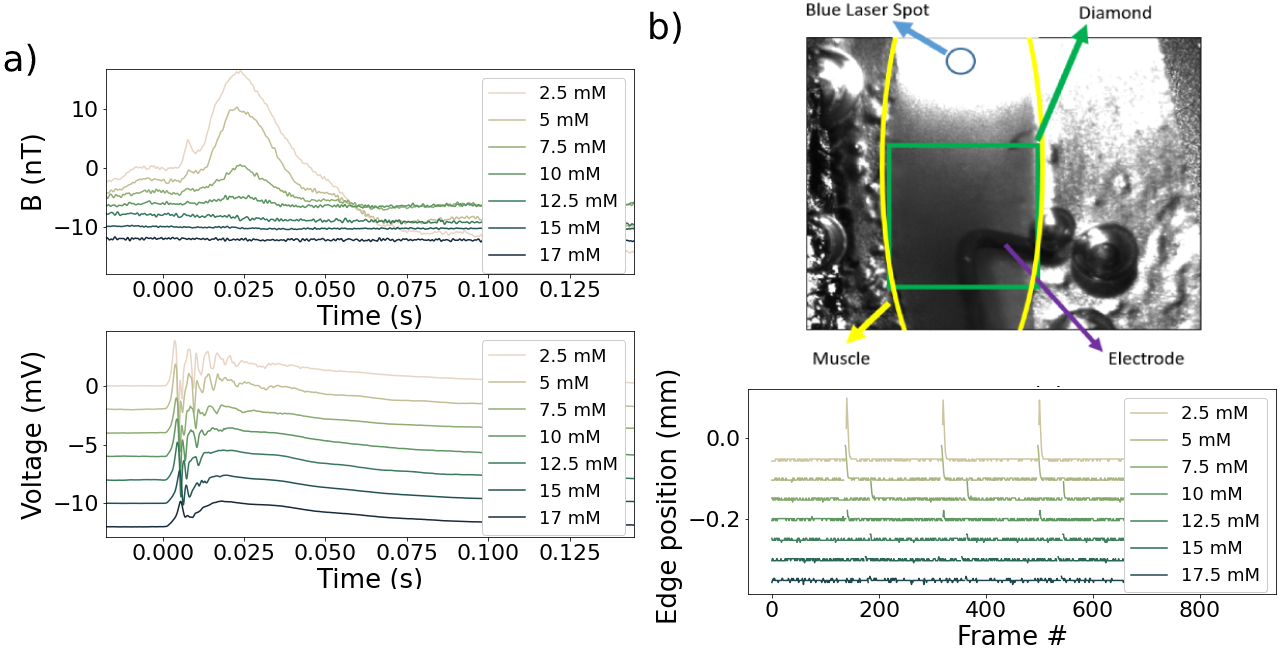}
\caption{\textbf{The diamond NV sensor and optical response to stimulation as a function of increasing BDM concentration.} a) NV sensor response to stimulation - with low concentrations of BDM a large artifact associated with muscle movement was observed; this was eliminated at concentrations above 15mM. Bottom: Electric data for the same muscle. Here, oscillations arising from movement of the electrode on the moving muscle are reduced in favour of the biosignal of the electrical activity in the muscle. b) To fully confirm the suppression of movement, we used white light microscopy, imaging the muscle edge during stimulation with a camera at 90fps. Above 15mM, spikes associated with muscle movement are completely eliminated.  }
\label{Fig3}
\end{figure} 

The strong contraction of the muscle after laser stimulation was found to produce a movement-induced artifact in our diamond NV sensor. This artifact arose due to a slight shift in microwave resonance frequency resulting from a motion-induced change in coupling between the nearfield antenna and the diamond and tissue above it. In order to remove this artifact, we pharmacologically suppressed movement using the myosin ATPase inhibitor 2,3-Butanedione monoxime (BDM), without affecting compound action potential propagation. To determine the necessary concentration, we gradually increased the concentration of BDM in the recycled ACSF perfused through the solution bath, starting at 2.5 mM and increasing in steps of 2.5 mM at 10 minute intervals. 

Figure \ref{Fig3},a) shows the measured NV sensor response (magnetic data) for a laser stimulated muscle at a range of different concentrations of BDM. With no or low concentration of BDM, we observed a significant movement artifact, equivalent to a 15nT magnetic field signal, followed by a decaying response lasting over 200ms. This artifact completely masked the sought action potential signal from the muscle. The motion artifact was considerably larger for laser stimulation than observed in our previous work using LED stimulation. We attribute this to significantly more effective stimulation of the muscle with the higher intensity laser than the weaker LED light, with the stronger movement clearly visible even by eye.  

By increasing the concentration of BDM from 2.5 to 20mM in the ACSF solution, we observed a clear reduction in the movement artifact, with complete elimination above 15mM. This could also be observed in the electric signal from the surface electrode, with oscillations reflecting the movement of the electrode on the muscle eliminated in favour of the electrical response in the muscle (Figure \ref{Fig3}, a). In order to confirm the absence of movement, we used a white light microscope and camera mounted above the sample, recording the movement of the muscle edge as a function of time while stimulating with the blue laser. With low concentrations of BDM, clear jumps in position could be observed post-stimulation as the muscle moved (Figure \ref{Fig3},b). Above 15mM of BDM, these were completely suppressed, with no movement detectable anywhere along the muscle. 

\subsection{Noise Filtering}

\begin{figure}[!ht]
\begin{centering}
\includegraphics[width=164mm]{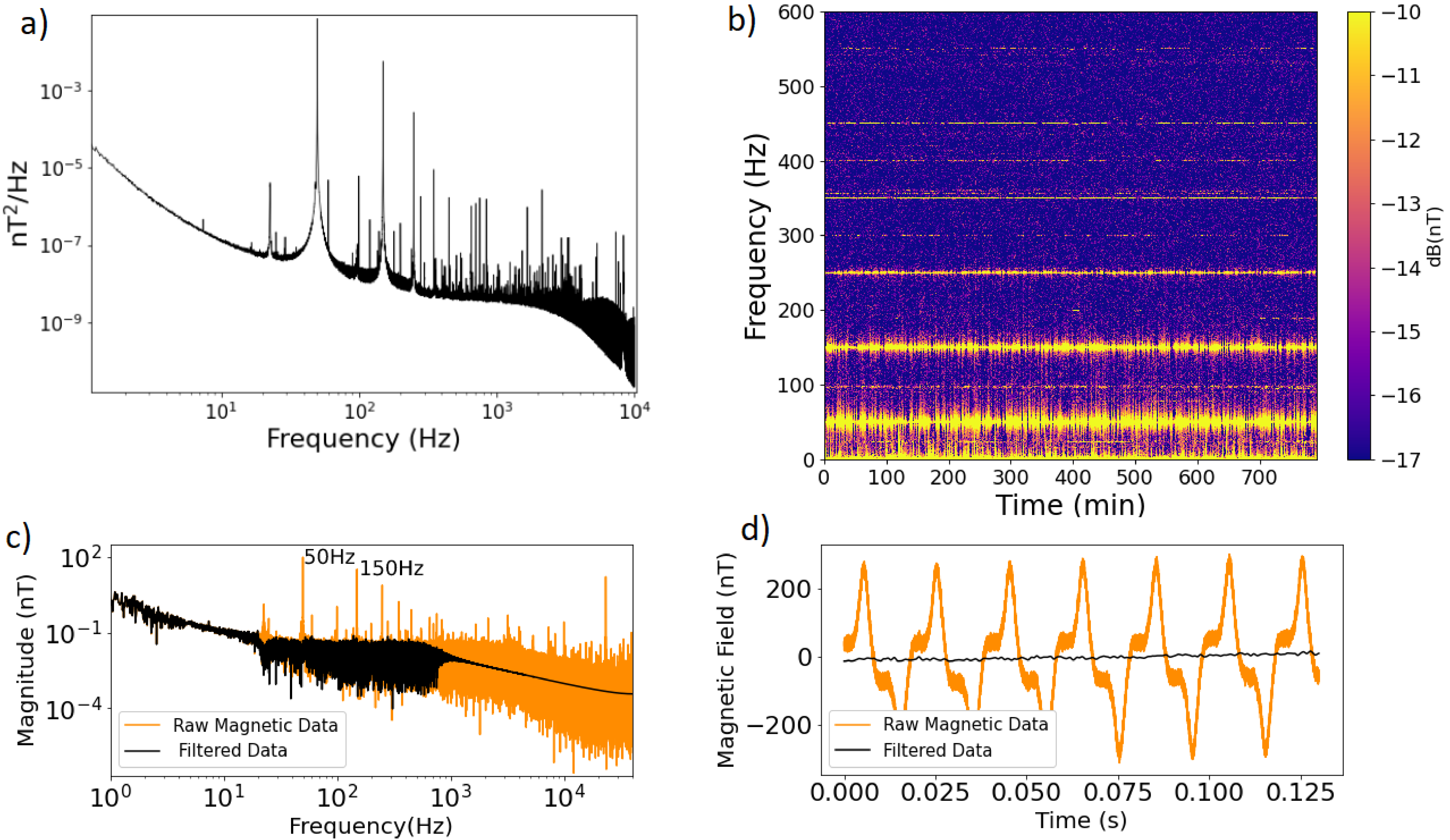}
\caption{\textbf{The noise spectrum of the diamond NV sensor, before and after post-processing filtering.} a) Averaged spectrum of the background magnetic noise as measured by the sensor and b) spectrogram of the same noise taken over 13 hours (normalised to 1nT). We observe significant noise arising from sources such as mains electricity in the $<$1kHz region that contains the majority of the components of the ms-scale biological signal. c) Comparison of the amplitude spectrum before and after the application of the whitening filter, with a significant amount of noise identified and removed. This is reflected in the example timeseries shown in d), showing a 0.5sec trace before and after the filter is implemented. Filtering reduces the noise from peak-to-peak of 400nT to less than 1nT on a single 60sec trace. }
\label{Fig4}
\end{centering}
\end{figure} 

In Figure \ref{Fig4},a) and b) we show the unfiltered power spectral density (PSD) from data recorded using our diamond NV sensor, containing significant noise from background magnetic field sources. We show the effect of our whitening filter algorithm in Figure \ref{Fig4},c) and b). As can be observed in the spectrum and the post-filtered timeseries, the filter significantly reduced the noise level. The PSD used to derive the filter was calculated using Welch's method with time intervals of 3.5s, 50\% overlap and Hanning window. In Supplementary Information, we outline the process by which we optimise the parameters of the whitening filter to remove noise, while retaining sufficient frequency components of the biosignal. We also directly compare the whitening filter performance against that derived using notch filtering and clarify how real absolute magnetic field units (Tesla) can be recovered from the relative PSD of the whitened data. Overall, we found the performance of the whitening filter in terms of noise removal was not significantly better than by using the notch filter method. However, it was significantly faster ($\times$2.5 speedup) and easier to implement, requiring only the calculation of the PSD for each 60s time trace compared to a serial algorithm needed to sweep through and build the notch filter mask. We found this could also be sped up by performing an initial time domain filter step via principle component analysis of the raw signal\citep{Negishi2004}(see SI). For the majority of biological applications, the signal shape and relative amplitude is of main concern, making the whitening method a fast and effective option.  

\subsection{Biosignal Readout}
\begin{figure}[!ht]
\begin{centering}
\includegraphics[width=164mm]{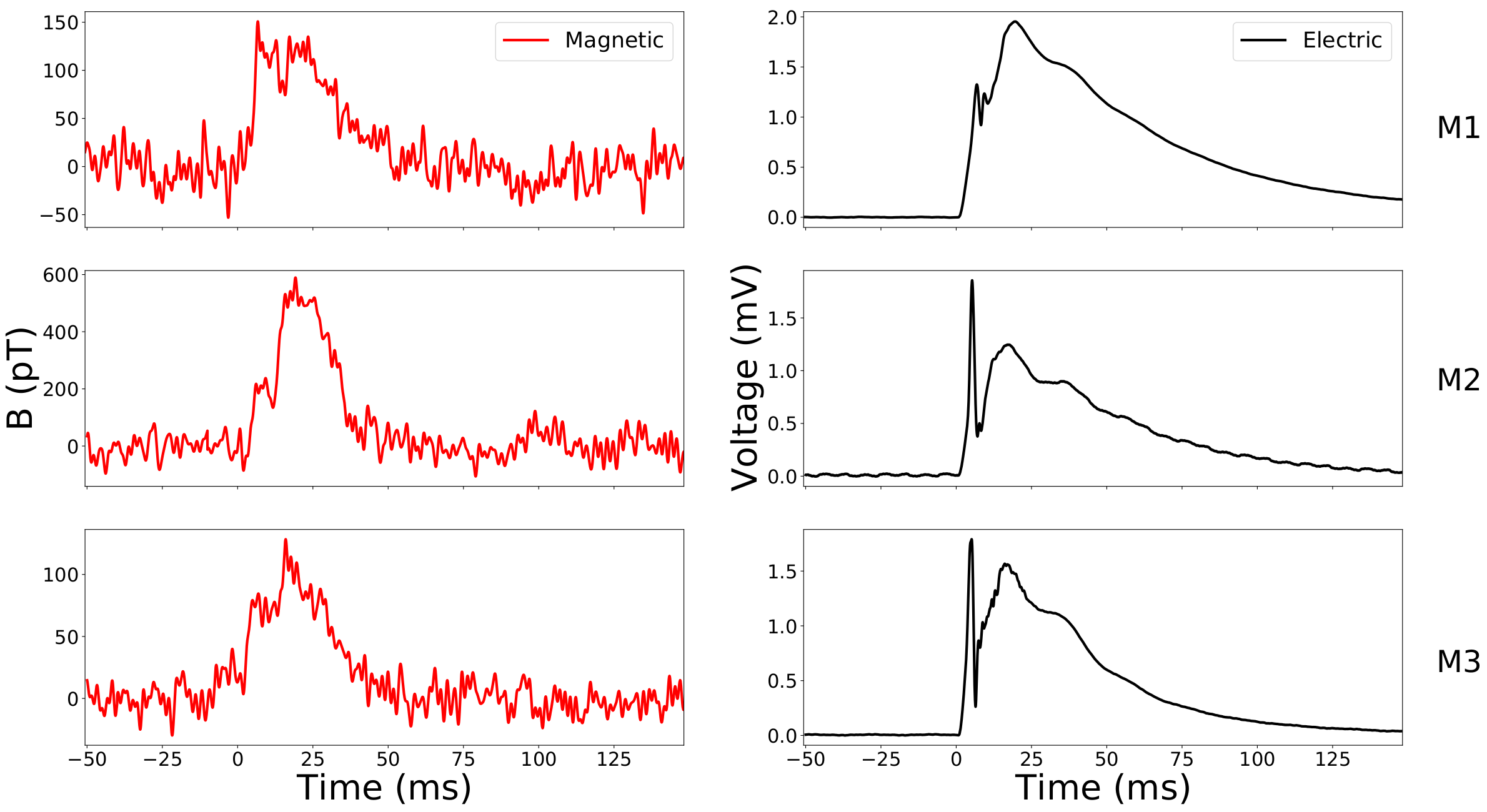}
\caption{\textbf{Measuring the muscle biological response electrically and magnetically}. Magnetic and electric data recorded for 3 different muscles (M1-M3). The data shown is the average response over 15000, 13500 and 540 stimulations for M1-3 respectively. The action potential (AP) peak and the channelrhodopsin (CH) are clearly distinguishable on all three muscles. On all subfigures, time t=0 corresponds to the instant laser stimulation is triggered.}
\label{Fig6}
\end{centering}
\end{figure} 

\begin{figure}[!ht]
\begin{centering}
\includegraphics[width=164mm]{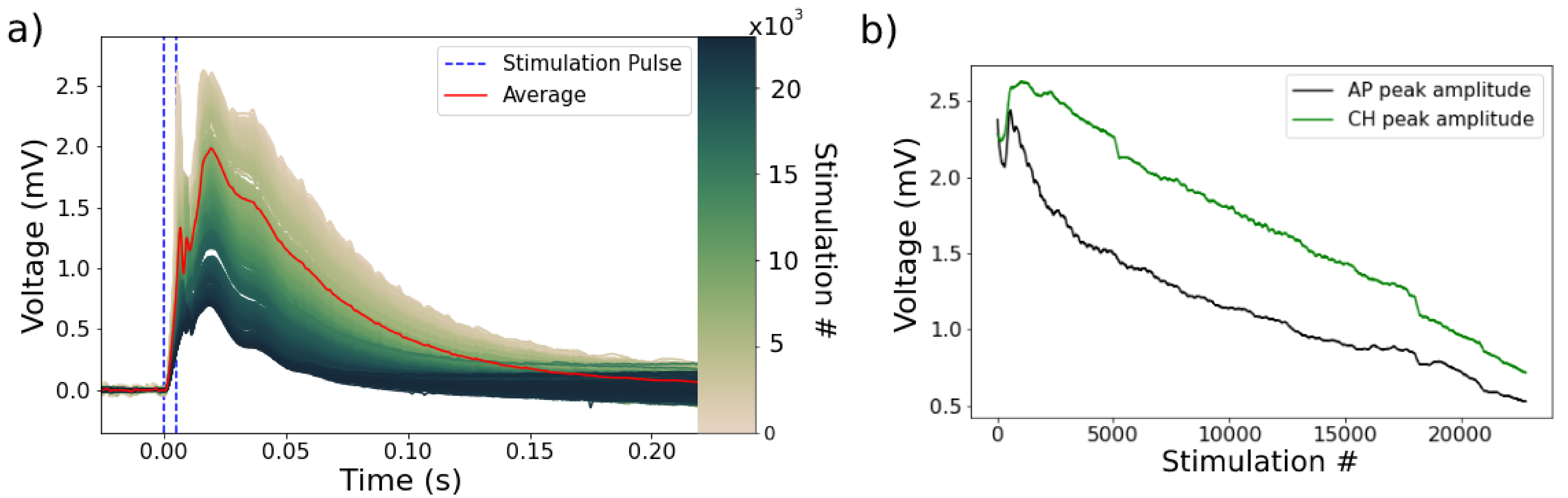}
\caption{\textbf{Stability of the measured biological response over time}. a) Variations in electric data recording versus number of stimulations for muscle M1. The amplitude of the signal gradually decreased over time as the muscle fatigued. We observed both a fast response peak from the action potential at $\sim$5ms and a second slower peak from channelrhodopsin at $\sim$20ms. b) Scaling of the standard deviation of the 60s timeseries (measuring noise) as a function of number of stimulations for both notch and whitening filtering, showing the decay of the AP and CH peaks.}
\label{Fig7}
\end{centering}
\end{figure} 
 
In Figure \ref{Fig6},a) we show examples of the biological signal measured by the diamond NV sensor (\textit{magnetic data}) and simultaneously recorded electrically using the AgCl contact electrode (\textit{electric data}) for 3 different muscles (M1-3), each recorded in a separate experimental run. For each muscle, we observed both a fast response peak (AP) at $\sim$5ms and a slower response (CH) at $\sim$10ms after the stimulation trigger. We attribute the first peak (AP) to the compound action potential response and the second slower peak (CH) to the response from channelrhodopsin activity. Both signals decayed in amplitude over many stimulations while retaining a similar shape, with the AP signal decaying more rapidly and begin subsumed into the CH signal after 4-6 hours of measurement (Figure \ref{Fig7}).  A slight delay in time was observed between the electric and magnetic data, arising from the positioning of the probe electrode away from the stimulation site, but still within 0.2-0.5 mm of the 2 $\times$2 mm$^2$ diamond sensor.  We use a compromise upper cutoff frequency of 650Hz, minimising distortion to both AP and CH signals (see SI), but acting to smooth the AP signal in the magnetic data. 
We note that the biological signal we observe using laser stimulation is significantly different from that previously observed for LED stimulation. The AP peak amplitude was weaker by a factor of 2 and in both the electric and magnetic data and recording shown in Figure \ref{Fig3}, a), we observed a non-negative refractory period followed by a slowly (tens of milliseconds) decaying response (CH). Furthermore, we observe a difference in the relative height of the AP and CH peaks between the magnetic and electrical readouts. We attribute this to the channelrhodopsin channels opening faster and remaining open for longer under higher intensity laser light, allowing ionic current to flow for longer and generating the strong CH signal. 

The peak magnetic field observed for muscle M2 was a factor of 5 stronger than for M1 and M3. We consider this likely to arise due to unexpected variability in sensor response across the 2 $\times$2 mm$^2$ diamond area. Due to the high level of internal reflection in high refractive index diamond, that we collect light from the full diamond volume and that the diamond had a relatively uniform density of NV centres, we expected sensor output to not vary significantly depending on muscle position, so long as the muscle was positioned approximately above the sensor. However, we found this not to be the case. Later investigations using a scanned electrified probe tip across the diamond surface, recording the magnetic field it induced, determined an area approximately 0.5$\times$0.5 mm to be a factor 16 better sensitivity than the lowest response elsewhere on the diamond (see SI). This region corresponded approximately to where the pump laser beam struck the diamond, receiving the maximum laser intensity. 

Although we cannot directly confirm the precise positioning, in the case of M2 it is likely the muscle was located optimally at the most sensitive region. For M1 and M2, the location was suboptimal but enough to yield a signal. For a number of other muscles, likely located away from this region, we recorded no biosignal in the magnetic data. We consider the positional response a particularly important result for this type of sensor, highlighting the need to correctly position the biosamples and spatially map sensor response. We also highlight that this potentially offers a novel route to spatially resolve and image signals well below the physical dimensions of the sensor using a scanned, highly focused pump laser beam. Further investigations of these aspects are called for, but are beyond the scope of this work.

\subsection{Simulation and Modeling} 
\begin{figure}[!ht]
\begin{centering}
\includegraphics[width=164mm]{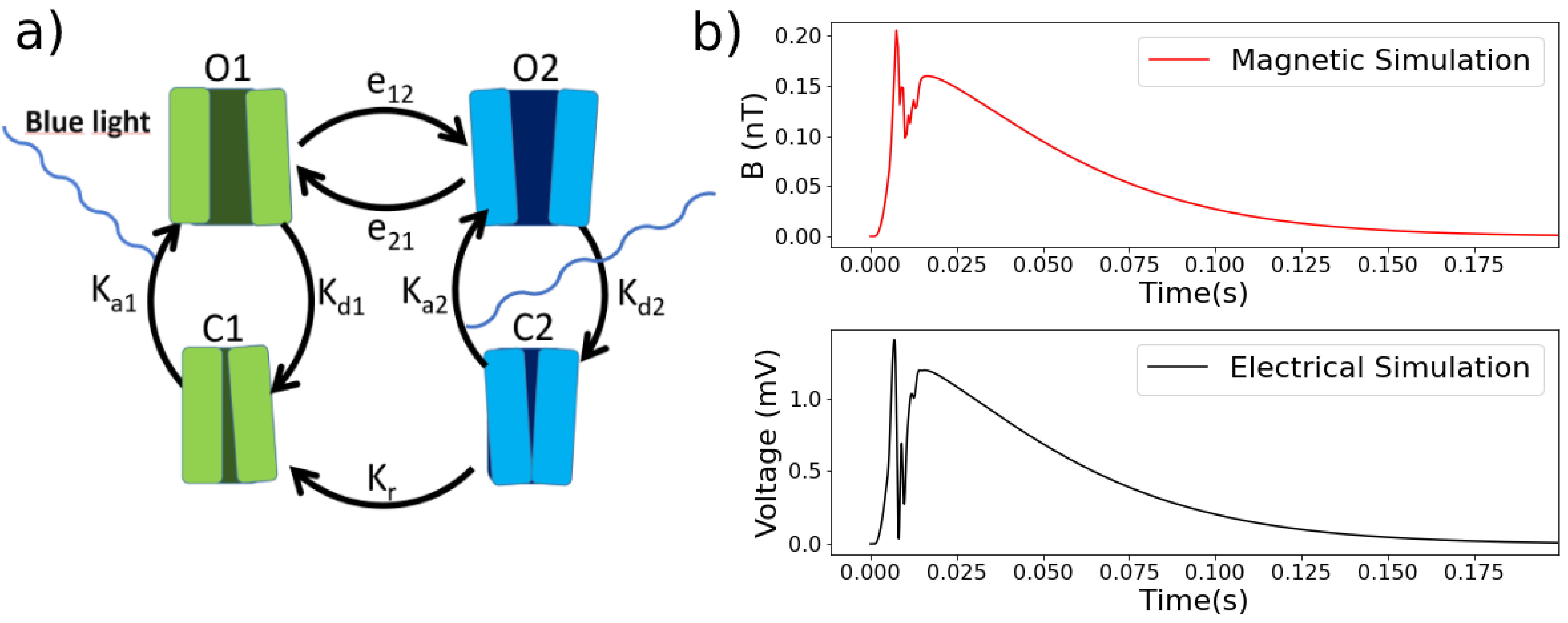}
\caption{\textbf{Channelrhodopsin model as related to our recorded response}. a) Depiction of the different stages of the four-state model of a ChR2 molecule. Blue light can transition the ChR2 from closed (C1 and C2) to open (O1 and O2). b) Upper figure show simulated magnetic signal as detected by averaging over the whole diamond. Lower figure show simulated LFP as detected by an electrode positioned as in Figure \ref{Fig1},c. Stimulation is modeled to occur at $t$=0.}
\label{Fig8}
\end{centering}
\end{figure}  

In order to explore the origin of the observed signal features, we modelled the muscle system to calculate the response under laser stimulation. In both electric and magnetic data, we observed a signal (CH) with a slow decay (Figure \ref{Fig6}). We attribute this signal to the activation of channelrhodopsin, with such a signal previously observed for optogenetically stimulated muscle signals\citep{Bruegmann2015, Magown2015}. When trying to model this response, we found that the original biophysical models of ChR2, developed for stimulation of neurons rather than muscle, could not describe the length of the CH signal we observe. Modelling the dynamics of ChR2 in neurons for a large span of different scenarios, including covering the range of current and intensity in this work, predict that any signal should decay within 15 ms of the stimulation \citep{Nikolic2009}. 

The four-state ChR2 model \citep{Grossman2011} for this process is shown in Figure \ref{Fig8},a). The ChR2-proteins are mainly populated at rest in the closed fast state (C1), which, when illuminated, transitions to the open fast state (O1) with transition rate K$_{a1}$ (stimulation laser intensity dependent) thus allowing influx of ions to the fiber. The O1 is unstable and decays mainly to C1 (with transition rate K$_{d1}$) as long as short excitation pulses ($\approx$5 ms) are used. This typically results in a single action potential without any pronounced secondary prolonged current.

With longer pulse duration, C1 can also transition to the slow opened state (O2). This transition is characterised with the irradiance dependent transition rates e$_{12}$ and e$_{21}$. The O2 state relaxes relatively slow to the closed slow state (C2) with a transition rate of K$_{d2}$, which in turn can either transition back to O2 given an excitation, with transition rate K$_{a2}$, or return to C1 with transition rate K$_{r}$. As the O2 state stays open for a longer time period after the stimulation pulse, it also creates a longer decay of the signal. This was experimentally observed for neuron stimulation for excitation pulses exceeding 50ms \citep{Nikolic2009}. 

However, we observe our slowly decaying CH signal at a shorter stimulation period of 5 ms. This suggests that the ChR2 in the studied muscle transitioned into the O2 state with larger proportion than in neurons. We therefore modified the four-state model to account for this behaviour (see SI for full details). Briefly, we modeled the induction of a secondary long-lasting current at a stimulation period of 5 ms by allowing for a stronger transition to the O2 configuration in ChR2, decaying more slowly to the closed state. This was done by setting the e$_{12}$ and e$_{21}$ parameters higher during illumination, and then letting e$_{12}$ remain high after illumination, whereas e$_{21}$ returned to a low value (Table SI2). Furthermore, the conductivity of the O2 state was also set higher than in previous models (from about 2-5fS to 25fS). These alternative values for the ChR2 model are justified by considering that the original ChR2 model was developed and tested for systems with lower stimulation intensities and for neuronal structures instead of muscle. Modification of these few parameters gives a modelled signal (Figure \ref{Fig8}, b) highly representative of the magnetic and electric data recorded experimentally (Figure \ref{Fig6}).

\section{Conclusions}

In this work, we demonstrate the viability of a diamond nitrogen-vacancy (NV) colour centre biosensor for recording biosignals while stimulating activity in high proximity to the sensor using focused pulsed laser light. Although we use simple muscle tissue, which could be easily probed electrically, our results show that highly localised, high intensity, selective excitation of specific biological processes in more complex systems can be done while passively sensing in the same localised region. This is essential for realising proposals such as to mapping neural activity in networks in the brain \citep{alivisatos2012} and experiments using intracellular nanodiamonds \citep{schirhagl2014}. 

We show that laser optogenetic stimulation of mouse muscle produces a stronger physical response, resulting in a significant movement artifact in the biosensor readout. We were able to fully suppress this artifact using a muscle movement inhibitor, although suppression was more challenging with higher intensity laser illumination than low intensity LED light. These results highlight the importance of understanding and eliminating unwanted artifacts in biosensing experiments. Although in this work the motion artifact represented a disadvantage, it could be exploited in a novel type of diamond NV sensor for force/motion detection, such as vector sensing of force/movement in a biological sample \citep{Cohen2020}. 

Using a filter algorithm based on spectral whitening we shown similar noise reduction performance as for using a notched filter, but with an improvement in terms of post-processing filter time ($\approx$2.5 times faster), especially where recovery of precise field units are not required. Processing speed could be further enhanced by an initial time domain filter step using principle component analysis to remove the main noise components (e.g. 50/150Hz inductive mains). This represents a step forward in terms of realising real-time biological signal recovery in an ordinary, unshielded lab environment  Further advances in filtering may be obtained by implementing artificial intelligence techniques in filter design, in particular by supplying synthetic known biosignals to the sensor. 

We observe both short period (AP) and long period (CH) signals in the magnetic data recorded by our NV sensor. Through modeling, we attribute these signals to muscle action potential (AP) and to current flow through channelrhodopsin (CH), with more muscle fibres addressed by the deeply penetrating laser beam and the channel opened for longer by the higher laser intensity. Although not ideal, with the channelrodopsin activation obscuring the action potential signal, the result exemplifies the usefulness of our diamond NV sensor in gaining new insight into a biological system close to the  optical stimulation site. Using alternative techniques, artifacts arising from photovoltaic effects (for electrical probes) or optical bleaching (for fluorescent biomarkers) would entirely mask the biosignal. The muscle position variability of the strength of our magnetic response highlights the need to properly characterise the sensor, and the potential to record strong signals if this process is well optimised. These results also potentially point to a way to achieve spatial resolution below the physical dimensions of the diamond, while still maintaining the same optical collection and excitation scheme. Such spatial resolution is key to fully, microscopically resolve and identify the origin of activity in living biological samples. 

\section{Competing Interests}
Hartwig R. Siebner has received honoraria as speaker from Sanofi Genzyme, Denmark and Novartis, Denmark, as consultant from Sanofi Genzyme, Denmark, Lophora, Denmark, and Lundbeck AS, Denmark, and as editor-in-chief (Neuroimage Clinical) and senior editor (NeuroImage) from Elsevier Publishers, Amsterdam, The Netherlands and royalties as book editor from Springer Publishers, Stuttgart, Germany and from Gyldendal Publishers, Copenhagen, Denmark. 

\section{Ethical Statement}
The work described in this article has been carried out in accordance with Directive 86/609/EEC for animal experiments and all relevant national legislation in Denmark. 

\section{Funding}
This work was funded by the Novo Nordisk foundation through the synergy grant bioQ (Grant Number: NNF17OC0028086) and the Center for Macroscale Quantum States (bigQ) funded by the Danish National Research Foundation (Grant number:DNRF142). Hartwig R. Siebner holds a 5-year professorship in precision medicine at the Faculty of Health Sciences and Medicine, University of Copenhagen funded by the Lundbeck Foundation (Grant Nr. R186). 

 \bibliographystyle{apa} 

\begin{thebibliography}{}

\bibitem[\protect\astroncite{Alivisatos et~al.}{2012}]{alivisatos2012}
Alivisatos, A.~P., Chun, M., Church, G.~M., Greenspan, R.~J., Roukes, M.~L.,
  and Yuste, R. (2012).
\newblock The brain activity map project and the challenge of functional
  connectomics.
\newblock {\em Neuron}, 74(6):970--974.

\bibitem[\protect\astroncite{Arlett et~al.}{2011}]{arlett2011}
Arlett, J., Myers, E., and Roukes, M. (2011).
\newblock Comparative advantages of mechanical biosensors.
\newblock {\em Nature nanotechnology}, 6(4):203--215.

\bibitem[\protect\astroncite{Augusto et~al.}{2017}]{Augusto2017}
Augusto, V., Padovani, C.~R., and Campos, G. E.~R. (2017).
\newblock {Skeletal muscle fiber types in C57BL6J mice}.
\newblock {\em Journal of Morphological Sciences}, 21(2):0.

\bibitem[\protect\astroncite{Barry et~al.}{2020}]{Barry2020}
Barry, J.~F., Schloss, J.~M., Bauch, E., Turner, M.~J., Hart, C.~A., Pham,
  L.~M., and Walsworth, R.~L. (2020).
\newblock Sensitivity optimization for {NV}-diamond magnetometry.
\newblock {\em Reviews of Modern Physics}, 92(1).

\bibitem[\protect\astroncite{Barry et~al.}{2016}]{Barry2016}
Barry, J.~F., Turner, M.~J., Schloss, J.~M., Glenn, D.~R., Song, Y., Lukin,
  M.~D., Park, H., and Walsworth, R.~L. (2016).
\newblock Optical magnetic detection of single-neuron action potentials using
  quantum defects in diamond.
\newblock {\em Proceedings of the National Academy of Sciences},
  113(49):14133--14138.

\bibitem[\protect\astroncite{Baselt et~al.}{1998}]{baselt1998}
Baselt, D.~R., Lee, G.~U., Natesan, M., Metzger, S.~W., Sheehan, P.~E., and
  Colton, R.~J. (1998).
\newblock A biosensor based on magnetoresistance technology.
\newblock {\em Biosensors and Bioelectronics}, 13(7-8):731--739.

\bibitem[\protect\astroncite{Bernardi et~al.}{2020}]{Bernardi2020}
Bernardi, E., Moreva, E., Traina, P., Petrini, G., Tchernij, S.~D., Forneris,
  J., Pastuovi{\'{c}}, {\v{Z}}., Degiovanni, I.~P., Olivero, P., and Genovese,
  M. (2020).
\newblock A biocompatible technique for magnetic field sensing at (sub)cellular
  scale using nitrogen-vacancy centers.
\newblock {\em {EPJ} Quantum Technology}, 7(1).

\bibitem[\protect\astroncite{Boto et~al.}{2018}]{Boto2018}
Boto, E., Holmes, N., Leggett, J., Roberts, G., Shah, V., Meyer, S.~S.,
  Mu{\~{n}}oz, L.~D., Mullinger, K.~J., Tierney, T.~M., Bestmann, S., Barnes,
  G.~R., Bowtell, R., and Brookes, M.~J. (2018).
\newblock Moving magnetoencephalography towards real-world applications with a
  wearable system.
\newblock {\em Nature}, 555(7698):657--661.

\bibitem[\protect\astroncite{Bruegmann et~al.}{2015}]{Bruegmann2015}
Bruegmann, T., van Bremen, T., Vogt, C.~C., Send, T., Fleischmann, B.~K., and
  Sasse, P. (2015).
\newblock Optogenetic control of contractile function in skeletal muscle.
\newblock {\em Nature Communications}, 6(1).

\bibitem[\protect\astroncite{Cannon et~al.}{1993}]{Cannon1993}
Cannon, S., Brown, R., and Corey, D. (1993).
\newblock Theoretical reconstruction of myotonia and paralysis caused by
  incomplete inactivation of sodium channels.
\newblock {\em Biophysical Journal}, 65(1):270--288.

\bibitem[\protect\astroncite{Cardin et~al.}{2010}]{Cardin2010-tr}
Cardin, J.~A., Carl{\'e}n, M., Meletis, K., Knoblich, U., Zhang, F.,
  Deisseroth, K., Tsai, L.-H., and Moore, C.~I. (2010).
\newblock Targeted optogenetic stimulation and recording of neurons in vivo
  using cell-type-specific expression of channelrhodopsin-2.
\newblock {\em Nat. Protoc.}, 5(2):247--254.

\bibitem[\protect\astroncite{Chatziioannou et~al.}{2019}]{Chatziioannou2019}
Chatziioannou, K., Haster, C.-J., Littenberg, T.~B., Farr, W.~M., Ghonge, S.,
  Millhouse, M., Clark, J.~A., and Cornish, N. (2019).
\newblock Noise spectral estimation methods and their impact on gravitational
  wave measurement of compact binary mergers.
\newblock {\em Physical Review D}, 100(10).

\bibitem[\protect\astroncite{Clarke et~al.}{2018}]{Clarke2018}
Clarke, J., Lee, Y.-H., and Schneiderman, J. (2018).
\newblock Focus on {SQUIDs} in biomagnetism.
\newblock {\em Superconductor Science and Technology}, 31(8):080201.

\bibitem[\protect\astroncite{Cohen et~al.}{2020}]{Cohen2020}
Cohen, D., Nigmatullin, R., Kenneth, O., Jelezko, F., Khodas, M., and Retzker,
  A. (2020).
\newblock Utilising {NV} based quantum sensing for velocimetry at the
  nanoscale.
\newblock {\em Scientific Reports}, 10(1).

\bibitem[\protect\astroncite{de~Cheveign{\'e} and
  Arzounian}{2018}]{de2018robust}
de~Cheveign{\'e}, A. and Arzounian, D. (2018).
\newblock Robust detrending, rereferencing, outlier detection, and inpainting
  for multichannel data.
\newblock {\em NeuroImage}, 172:903--912.

\bibitem[\protect\astroncite{Dolde et~al.}{2011}]{Dolde2011}
Dolde, F., Fedder, H., Doherty, M.~W., N\"{o}bauer, T., Rempp, F.,
  Balasubramanian, G., Wolf, T., Reinhard, F., Hollenberg, L. C.~L., Jelezko,
  F., and Wrachtrup, J. (2011).
\newblock Electric-field sensing using single diamond spins.
\newblock {\em Nature Physics}, 7(6):459--463.

\bibitem[\protect\astroncite{El-Ella et~al.}{2017}]{ElElla2017}
El-Ella, H. A.~R., Ahmadi, S., Wojciechowski, A.~M., Huck, A., and Andersen,
  U.~L. (2017).
\newblock Optimised frequency modulation for continuous-wave optical magnetic
  resonance sensing using nitrogen-vacancy ensembles.
\newblock {\em Optics Express}, 25(13):14809.

\bibitem[\protect\astroncite{Fescenko et~al.}{2020}]{Fescenko2020}
Fescenko, I., Jarmola, A., Savukov, I., Kehayias, P., Smits, J., Damron, J.,
  Ristoff, N., Mosavian, N., and Acosta, V.~M. (2020).
\newblock Diamond magnetometer enhanced by ferrite flux concentrators.
\newblock {\em Physical Review Research}, 2(2).

\bibitem[\protect\astroncite{Foutz et~al.}{2012}]{Foutz2012}
Foutz, T.~J., Arlow, R.~L., and McIntyre, C.~C. (2012).
\newblock Theoretical principles underlying optical stimulation of a
  channelrhodopsin-2 positive pyramidal neuron.
\newblock {\em Journal of Neurophysiology}, 107(12):3235--3245.

\bibitem[\protect\astroncite{Fujiwara et~al.}{2020}]{Fujiwara2020}
Fujiwara, M., Sun, S., Dohms, A., Nishimura, Y., Suto, K., Takezawa, Y.,
  Oshimi, K., Zhao, L., Sadzak, N., Umehara, Y., Teki, Y., Komatsu, N., Benson,
  O., Shikano, Y., and Kage-Nakadai, E. (2020).
\newblock Real-time nanodiamond thermometry probing in vivo thermogenic
  responses.
\newblock {\em Science Advances}, 6(37):eaba9636.

\bibitem[\protect\astroncite{Gabbard et~al.}{2018}]{Gabbard2018}
Gabbard, H., Williams, M., Hayes, F., and Messenger, C. (2018).
\newblock Matching matched filtering with deep networks for gravitational-wave
  astronomy.
\newblock {\em Physical Review Letters}, 120(14).

\bibitem[\protect\astroncite{Gechev et~al.}{2016}]{GECHEV201662}
Gechev, A., Kane, N., Koltzenburg, M., Rao, D., and {van der Star}, R. (2016).
\newblock Potential risks of iatrogenic complications of nerve conduction
  studies (ncs) and electromyography (emg).
\newblock {\em Clinical Neurophysiology Practice}, 1:62--66.

\bibitem[\protect\astroncite{Grienberger and
  Konnerth}{2012}]{GRIENBERGER2012862}
Grienberger, C. and Konnerth, A. (2012).
\newblock Imaging calcium in neurons.
\newblock {\em Neuron}, 73(5):862--885.

\bibitem[\protect\astroncite{Grossman et~al.}{2011}]{Grossman2011}
Grossman, N., Nikolic, K., Toumazou, C., and Degenaar, P. (2011).
\newblock {Modeling study of the light stimulation of a neuron cell with
  channelrhodopsin-2 mutants}.
\newblock {\em IEEE Transactions on Biomedical Engineering}, 58(6):1742--1751.

\bibitem[\protect\astroncite{Gruber
  et~al.}{1997}]{doi:10.1126/science.276.5321.2012}
Gruber, A., DrÃ¤benstedt, A., Tietz, C., Fleury, L., Wrachtrup, J., and von
  Borczyskowski, C. (1997).
\newblock Scanning confocal optical microscopy and magnetic resonance on single
  defect centers.
\newblock {\em Science}, 276(5321):2012--2014.

\bibitem[\protect\astroncite{Hall et~al.}{2012}]{Hall2012}
Hall, L.~T., Beart, G. C.~G., Thomas, E.~A., Simpson, D.~A., McGuinness, L.~P.,
  Cole, J.~H., Manton, J.~H., Scholten, R.~E., Jelezko, F., Wrachtrup, J.,
  Petrou, S., and Hollenberg, L. C.~L. (2012).
\newblock High spatial and temporal resolution wide-field imaging of neuron
  activity using quantum {NV}-diamond.
\newblock {\em Scientific Reports}, 2(1).

\bibitem[\protect\astroncite{H\"{a}m\"{a}l\"{a}inen et~al.}{1993}]{Hmlinen1993}
H\"{a}m\"{a}l\"{a}inen, M., Hari, R., Ilmoniemi, R.~J., Knuutila, J., and
  Lounasmaa, O.~V. (1993).
\newblock Magnetoencephalography{\textemdash}theory, instrumentation, and
  applications to noninvasive studies of the working human brain.
\newblock {\em Reviews of Modern Physics}, 65(2):413--497.

\bibitem[\protect\astroncite{Hines and Carnevale}{2001}]{Hines2001}
Hines, M.~L. and Carnevale, N.~T. (2001).
\newblock Neuron: A tool for neuroscientists.
\newblock {\em The Neuroscientist}, 7(2):123--135.

\bibitem[\protect\astroncite{Horsley et~al.}{2018}]{PhysRevApplied.10.044039}
Horsley, A., Appel, P., Wolters, J., Achard, J., Tallaire, A., Maletinsky, P.,
  and Treutlein, P. (2018).
\newblock Microwave device characterization using a widefield diamond
  microscope.
\newblock {\em Phys. Rev. Applied}, 10:044039.

\bibitem[\protect\astroncite{Karadas et~al.}{2018}]{Karadas2018}
Karadas, M., Wojciechowski, A.~M., Huck, A., Dalby, N.~O., Andersen, U.~L., and
  Thielscher, A. (2018).
\newblock Feasibility and resolution limits of opto-magnetic imaging of neural
  network activity in brain slices using color centers in diamond.
\newblock {\em Scientific Reports}, 8(1).

\bibitem[\protect\astroncite{Kehayias et~al.}{2019}]{Kehayias2019}
Kehayias, P., Turner, M.~J., Trubko, R., Schloss, J.~M., Hart, C.~A., Wesson,
  M., Glenn, D.~R., and Walsworth, R.~L. (2019).
\newblock Imaging crystal stress in diamond using ensembles of nitrogen-vacancy
  centers.
\newblock {\em Physical Review B}, 100(17).

\bibitem[\protect\astroncite{Kozai and Vazquez}{2015}]{Kozai2015-om}
Kozai, T. D.~Y. and Vazquez, A.~L. (2015).
\newblock Photoelectric artefact from optogenetics and imaging on
  microelectrodes and bioelectronics: new challenges and opportunities.
\newblock {\em J. Mater. Chem. B Mater. Biol. Med.}, 3(25):4965--4978.

\bibitem[\protect\astroncite{Kucsko et~al.}{2013}]{Kucsko2013}
Kucsko, G., Maurer, P.~C., Yao, N.~Y., Kubo, M., Noh, H.~J., Lo, P.~K., Park,
  H., and Lukin, M.~D. (2013).
\newblock Nanometre-scale thermometry in a living cell.
\newblock {\em Nature}, 500(7460):54--58.

\bibitem[\protect\astroncite{Lin and Schnitzer}{2016}]{Lin2016-us}
Lin, M.~Z. and Schnitzer, M.~J. (2016).
\newblock Genetically encoded indicators of neuronal activity.
\newblock {\em Nat. Neurosci.}, 19(9):1142--1153.

\bibitem[\protect\astroncite{Lind{\'{e}}n et~al.}{2014}]{Linden2014}
Lind{\'{e}}n, H., Hagen, E., Leski, S., Norheim, E.~S., Pettersen, K.~H., and
  Einevoll, G.~T. (2014).
\newblock {LFPy: a tool for biophysical simulation of extracellular potentials
  generated by detailed model neurons}.
\newblock {\em Frontiers in neuroinformatics}, 7:41.

\bibitem[\protect\astroncite{Magown et~al.}{2015}]{Magown2015}
Magown, P., Shettar, B., Zhang, Y., and Rafuse, V.~F. (2015).
\newblock Direct optical activation of skeletal muscle fibres efficiently
  controls muscle contraction and attenuates denervation atrophy.
\newblock {\em Nature Communications}, 6(1).

\bibitem[\protect\astroncite{Mahajan et~al.}{2020}]{10.3389/fbioe.2020.00416}
Mahajan, S., Hermann, J.~K., Bedell, H.~W., Sharkins, J.~A., Chen, L., Chen,
  K., Meade, S.~M., Smith, C.~S., Rayyan, J., Feng, H., Kim, Y., Schiefer,
  M.~A., Taylor, D.~M., Capadona, J.~R., and Ereifej, E.~S. (2020).
\newblock Toward standardization of electrophysiology and computational tissue
  strain in rodent intracortical microelectrode models.
\newblock {\em Frontiers in Bioengineering and Biotechnology}, 8:416.

\bibitem[\protect\astroncite{McDonald}{2012}]{MCDONALD2012495}
McDonald, C.~M. (2012).
\newblock Clinical approach to the diagnostic evaluation of hereditary and
  acquired neuromuscular diseases.
\newblock {\em Physical Medicine and Rehabilitation Clinics of North America},
  23(3):495--563.
\newblock Neuromuscular Disease Management and Rehabilitation, Part I:
  Diagnostic and Therapy Issues.

\bibitem[\protect\astroncite{McGuinness et~al.}{2011}]{McGuinness2011}
McGuinness, L.~P., Yan, Y., Stacey, A., Simpson, D.~A., Hall, L.~T., Maclaurin,
  D., Prawer, S., Mulvaney, P., Wrachtrup, J., Caruso, F., Scholten, R.~E., and
  Hollenberg, L. C.~L. (2011).
\newblock Quantum measurement and orientation tracking of fluorescent
  nanodiamonds inside living cells.
\newblock {\em Nature Nanotechnology}, 6(6):358--363.

\bibitem[\protect\astroncite{Mizuno et~al.}{2020}]{Mizuno2020}
Mizuno, K., Ishiwata, H., Masuyama, Y., Iwasaki, T., and Hatano, M. (2020).
\newblock Simultaneous wide-field imaging of phase and magnitude of {AC}
  magnetic signal using diamond quantum magnetometry.
\newblock {\em Scientific Reports}, 10(1).

\bibitem[\protect\astroncite{Mr{\'{o}}zek et~al.}{2015}]{Mrzek2015}
Mr{\'{o}}zek, M., Rudnicki, D., Kehayias, P., Jarmola, A., Budker, D., and
  Gawlik, W. (2015).
\newblock Longitudinal spin relaxation in nitrogen-vacancy ensembles in
  diamond.
\newblock {\em {EPJ} Quantum Technology}, 2(1).

\bibitem[\protect\astroncite{Negishi et~al.}{2004}]{Negishi2004}
Negishi, M., Abildgaard, M., Nixon, T., and Constable, R.~T. (2004).
\newblock Removal of time-varying gradient artifacts from {EEG} data acquired
  during continuous {fMRI}.
\newblock {\em Clinical Neurophysiology}, 115(9):2181--2192.

\bibitem[\protect\astroncite{Neumann et~al.}{2013}]{Neumann2013}
Neumann, P., Jakobi, I., Dolde, F., Burk, C., Reuter, R., Waldherr, G., Honert,
  J., Wolf, T., Brunner, A., Shim, J.~H., Suter, D., Sumiya, H., Isoya, J., and
  Wrachtrup, J. (2013).
\newblock High-precision nanoscale temperature sensing using single defects in
  diamond.
\newblock {\em Nano Letters}, 13(6):2738--2742.

\bibitem[\protect\astroncite{Neumann et~al.}{2009}]{odmr2009}
Neumann, P., Kolesov, R., Jacques, V., Beck, J., Tisler, J., Batalov, A.,
  Rogers, L., Manson, N.~B., Balasubramanian, G., Jelezko, F., and Wrachtrup,
  J. (2009).
\newblock Excited-state spectroscopy of single {NV} defects in diamond using
  optically detected magnetic resonance.
\newblock {\em New Journal of Physics}, 11(1):013017.

\bibitem[\protect\astroncite{Nikolic et~al.}{2009}]{Nikolic2009}
Nikolic, K., Grossman, N., Grubb, M.~S., Burrone, J., Toumazou, C., and
  Degenaar, P. (2009).
\newblock Photocycles of channelrhodopsin-2.
\newblock {\em Photochemistry and Photobiology}, 85(1):400--411.

\bibitem[\protect\astroncite{Packer et~al.}{2013}]{Packer2013-vd}
Packer, A.~M., Roska, B., and H{\"a}usser, M. (2013).
\newblock Targeting neurons and photons for optogenetics.
\newblock {\em Nat. Neurosci.}, 16(7):805--815.

\bibitem[\protect\astroncite{Price et~al.}{2020}]{Price2020}
Price, J.~C., Mesquita-Ribeiro, R., Dajas-Bailador, F., and Mather, M.~L.
  (2020).
\newblock Widefield, spatiotemporal mapping of spontaneous activity of mouse
  cultured neuronal networks using quantum diamond sensors.
\newblock {\em Frontiers in Physics}, 8.

\bibitem[\protect\astroncite{Roman et~al.}{2000}]{Roman2000}
Roman, J., Rangaswamy, M., Davis, D., Zhang, Q., Himed, B., and Michels, J.
  (2000).
\newblock Parametric adaptive matched filter for airborne radar applications.
\newblock {\em {IEEE} Transactions on Aerospace and Electronic Systems},
  36(2):677--692.

\bibitem[\protect\astroncite{Sage et~al.}{2013}]{LeSage2013}
Sage, D.~L., Arai, K., Glenn, D.~R., DeVience, S.~J., Pham, L.~M., Rahn-Lee,
  L., Lukin, M.~D., Yacoby, A., Komeili, A., and Walsworth, R.~L. (2013).
\newblock Optical magnetic imaging of living cells.
\newblock {\em Nature}, 496(7446):486--489.

\bibitem[\protect\astroncite{Scanziani and HÃ¤usser}{2009}]{Scanzani2009}
Scanziani, M. and HÃ¤usser, M. (2009).
\newblock Electrophysiology in the age of light.
\newblock {\em Nature}, 461:930--939.

\bibitem[\protect\astroncite{Schirhagl et~al.}{2014}]{schirhagl2014}
Schirhagl, R., Chang, K., Loretz, M., and Degen, C.~L. (2014).
\newblock Nitrogen-vacancy centers in diamond: nanoscale sensors for physics
  and biology.
\newblock {\em Annual review of physical chemistry}, 65:83--105.

\bibitem[\protect\astroncite{Staudacher et~al.}{2013}]{Staudacher2013-ye}
Staudacher, T., Shi, F., Pezzagna, S., Meijer, J., Du, J., Meriles, C.~A.,
  Reinhard, F., and Wrachtrup, J. (2013).
\newblock Nuclear magnetic resonance spectroscopy on a (5-nanometer)3 sample
  volume.
\newblock {\em Science}, 339(6119):561--563.

\bibitem[\protect\astroncite{Steinert et~al.}{2010}]{doi:10.1063/1.3385689}
Steinert, S., Dolde, F., Neumann, P., Aird, A., Naydenov, B., Balasubramanian,
  G., Jelezko, F., and Wrachtrup, J. (2010).
\newblock High sensitivity magnetic imaging using an array of spins in diamond.
\newblock {\em Review of Scientific Instruments}, 81(4):043705.

\bibitem[\protect\astroncite{Taylor et~al.}{2008}]{Taylor2008}
Taylor, J.~M., Cappellaro, P., Childress, L., Jiang, L., Budker, D., Hemmer,
  P.~R., Yacoby, A., Walsworth, R., and Lukin, M.~D. (2008).
\newblock High-sensitivity diamond magnetometer with nanoscale resolution.
\newblock {\em Nature Physics}, 4(10):810--816.

\bibitem[\protect\astroncite{Toyli et~al.}{2012}]{toyli2012}
Toyli, D., Christle, D., Alkauskas, A., Buckley, B., Van~de Walle, C., and
  Awschalom, D. (2012).
\newblock Measurement and control of single nitrogen-vacancy center spins above
  600 k.
\newblock {\em Physical Review X}, 2(3):031001.

\bibitem[\protect\astroncite{Webb et~al.}{2020}]{Webb2020}
Webb, J.~L., Troise, L., Hansen, N.~W., Achard, J., Brinza, O., Staacke, R.,
  Kieschnick, M., Meijer, J., Perrier, J.-F., Berg-S{\o}rensen, K., Huck, A.,
  and Andersen, U.~L. (2020).
\newblock Optimization of a diamond nitrogen vacancy centre magnetometer for
  sensing of biological signals.
\newblock {\em Frontiers in Physics}, 8.

\bibitem[\protect\astroncite{Webb et~al.}{2021}]{Webb2021}
Webb, J.~L., Troise, L., Hansen, N.~W., Olsson, C., Wojciechowski, A.~M.,
  Achard, J., Brinza, O., Staacke, R., Kieschnick, M., Meijer, J., Thielscher,
  A., Perrier, J.-F., Berg-S{\o}rensen, K., Huck, A., and Andersen, U.~L.
  (2021).
\newblock Detection of biological signals from a live mammalian muscle using an
  early stage diamond quantum sensor.
\newblock {\em Scientific Reports}, 11(1).

\bibitem[\protect\astroncite{Wolf et~al.}{2015}]{Wolf2015}
Wolf, T., Neumann, P., Nakamura, K., Sumiya, H., Ohshima, T., Isoya, J., and
  Wrachtrup, J. (2015).
\newblock Subpicotesla diamond magnetometry.
\newblock {\em Physical Review X}, 5(4).

\bibitem[\protect\astroncite{Wu et~al.}{2021}]{Wu2021-ee}
Wu, Y., Alam, M. N.~A., Balasubramanian, P., Ermakova, A., Fischer, S., Barth,
  H., Wagner, M., Raabe, M., Jelezko, F., and Weil, T. (2021).
\newblock Nanodiamond theranostic for light-controlled intracellular heating
  and nanoscale temperature sensing.
\newblock {\em Nano Lett.}, 21(9):3780--3788.

\bibitem[\protect\astroncite{Xu et~al.}{2021}]{Xu2021-ab}
Xu, J., Jarocha, L.~E., Zollitsch, T., Konowalczyk, M., Henbest, K.~B.,
  Richert, S., Golesworthy, M.~J., Schmidt, J., D{\'e}jean, V., Sowood, D.
  J.~C., Bassetto, M., Luo, J., Walton, J.~R., Fleming, J., Wei, Y., Pitcher,
  T.~L., Moise, G., Herrmann, M., Yin, H., Wu, H., Bart{\"o}lke, R.,
  K{\"a}sehagen, S.~J., Horst, S., Dautaj, G., Murton, P. D.~F., Gehrckens,
  A.~S., Chelliah, Y., Takahashi, J.~S., Koch, K.-W., Weber, S., Solov'yov,
  I.~A., Xie, C., Mackenzie, S.~R., Timmel, C.~R., Mouritsen, H., and Hore,
  P.~J. (2021).
\newblock Magnetic sensitivity of cryptochrome 4 from a migratory songbird.
\newblock {\em Nature}, 594(7864):535--540.

\end{thebibliography}





\end{document}